\documentclass[twoside,leqno,twocolumn]{article}

\usepackage[letterpaper]{geometry}

\usepackage{ltexpprt}
\usepackage{hyperref}
\usepackage{marvosym} 
\usepackage{makecell}
\usepackage{amsmath}
\usepackage{graphicx}
\usepackage{float}
\usepackage{subfig}
\usepackage{overpic}
\usepackage{amsfonts}
\usepackage{caption}
\usepackage{amssymb}
\makeatletter
\newcommand{\rmnum}[1]{\romannumeral #1}
\newcommand{\Rmnum}[1]{\expandafter\@slowromancap\romannumeral #1@}
\makeatother
\usepackage{booktabs}
\usepackage{multirow}
 
\usepackage{stfloats}
\begin{document}

\newcommand\relatedversion{}
\newcommand\keywords[1]{\textbf{Keywords}: #1}




%
\title{\Large  Denoising Long- and Short-term Interests for Sequential Recommendation
  \relatedversion}

\author{
Xinyu Zhang \footnotemark[1]  \footnotemark[4] 
\and Beibei Li \footnotemark[2]  \footnotemark[4] 
\and Beihong Jin \footnotemark[3]  \textsuperscript{\Letter} }

\date{}

\maketitle
\renewcommand{\thefootnote}{\fnsymbol{footnote}}  
\footnotetext[1]{Institute of Software, Chinese Academy of Sciences. XinyuZhang0429@outlook.com.} 
\footnotetext[2]{College of Computer Science, Chongqing University. libeibei22@cqu.edu.cn.} 
\footnotetext[4]{These authors are co-first authors.} 
\footnotetext[3]{Corresponding author. Institute of Software, Chinese Academy of Sciences. University of Chinese Academy of Sciences. Beihong@iscas.ac.cn.}

\fancyfoot[R]{\scriptsize{Copyright \textcopyright\ 2024 by SIAM\\
Unauthorized reproduction of this article is prohibited}}

 \begin{abstract} \small\baselineskip=9pt User interests can be viewed over different time scales, mainly including stable long-term preferences and changing short-term intentions, and their combination facilitates the comprehensive sequential recommendation. However, existing work that focuses on different time scales of user modeling has ignored the negative effects of \textit{different time-scale noise},  which hinders capturing actual user interests and cannot be resolved by conventional sequential denoising methods. In this paper, we propose a \textbf{L}ong- and \textbf{S}hort-term \textbf{I}nterest \textbf{D}enoising \textbf{N}etwork (LSIDN), which employs different encoders and tailored denoising strategies to extract long- and short-term interests, respectively, achieving both comprehensive and robust user modeling. Specifically, we employ a session-level interest extraction and evolution strategy to avoid introducing \textit{inter-session behavioral noise} into long-term interest modeling; we also adopt contrastive learning equipped with a homogeneous exchanging augmentation to alleviate the impact of \textit{unintentional behavioral noise} on short-term interest modeling. Results of experiments on two public datasets show that LSIDN consistently outperforms state-of-the-art models and achieves significant robustness. 
 \end{abstract}

\noindent \keywords{recommender systems, sequential recommendation, interest modeling.}

\section{Introduction} 
Sequential Recommendation (SR)  \cite{fang2020deep} has been widely used in various platforms to alleviate information overload and provide personalized services for users. The task of SR is to predict the next-item that a user will be interested in, given his/her behavior sequence. In real-world scenarios, the user's interests can be viewed over different time scales, mainly including relatively stable long-term user preferences and more variable short-term user intentions, each of which reflects the different time-scale dynamics of the user historical sequence. Some work \cite{ying2018sequential,lv2019sdm,yu2019adaptive,tan2021dynamic,zheng2022disentangling,du2023idnp} has modeled both long- and short-term user interests separately and then combined them to facilitate more comprehensive user modeling, thereby improving recommendation performance.

\begin{figure*} 
         \centering
 \vspace{-1.5em}
    \begin{minipage}[t]{0.52\textwidth}
        \centering
        \includegraphics[width=\textwidth]{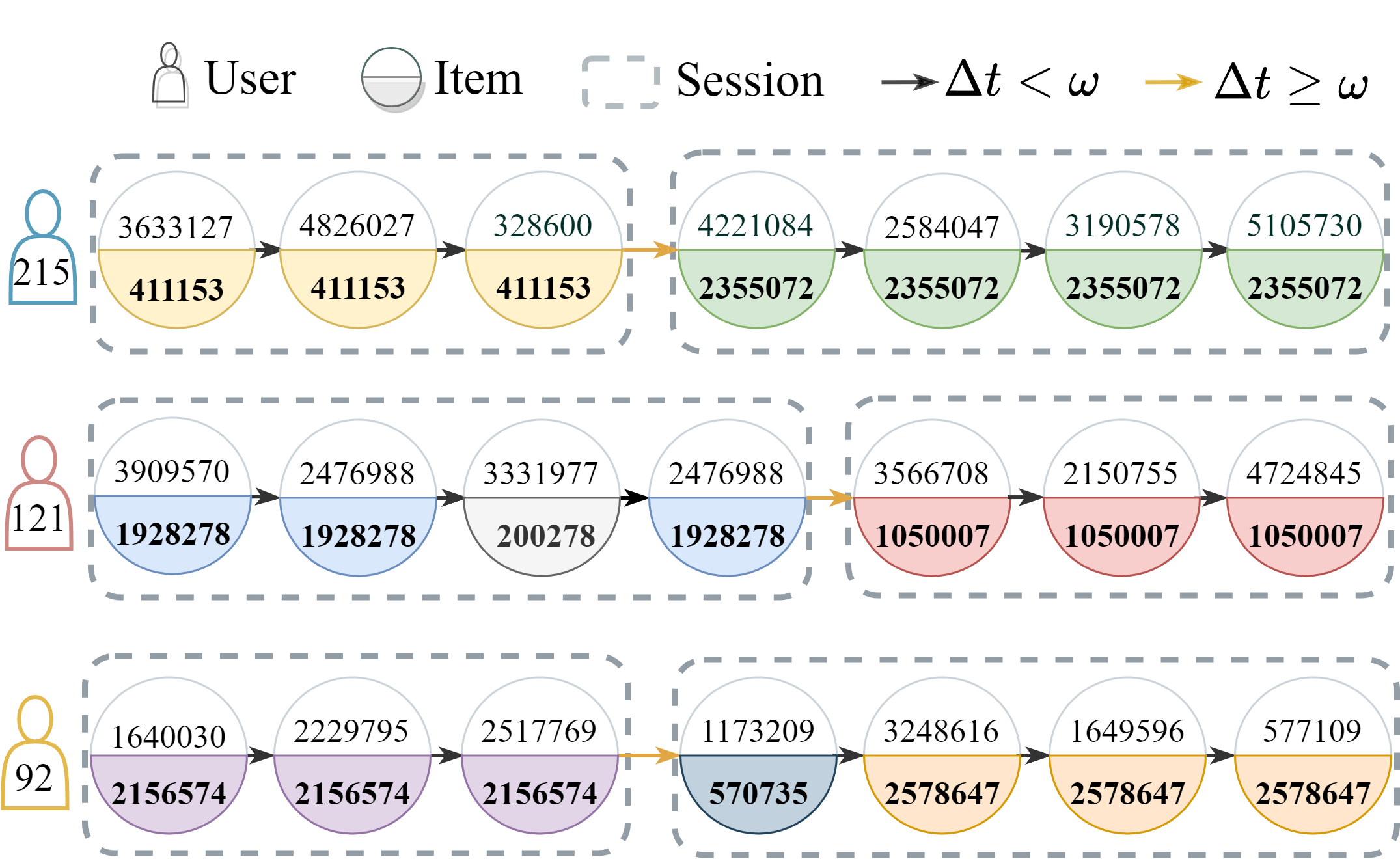 }
        \centerline{(a)}
    \end{minipage}%
    \hspace{.5in}
    \begin{minipage}[t]{0.256\textwidth}
        \centering
        \includegraphics[width=\textwidth]{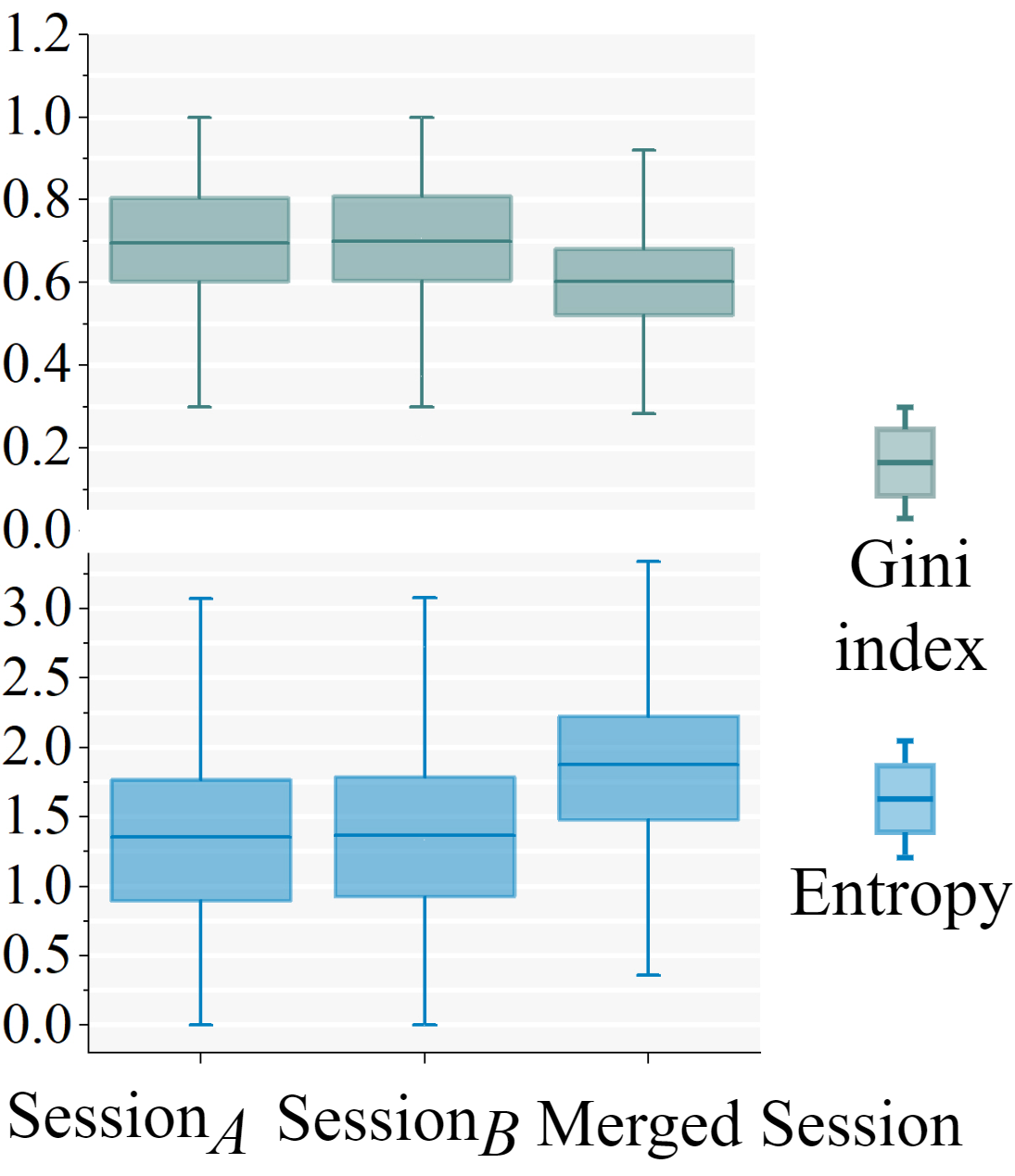}
        \centerline{(b)}
    \end{minipage}
    \caption{ Taking the Taobao dataset as an example, we divide the sequence into sessions according to the time interval $\Delta t$ between two successive items, and set the division threshold $\omega$ = 360min. Note that the diversity of item categories in a session/sequence can be used to reflect the diversity of user behaviors. (a) We randomly selected several sequence instances for qualitative observation. Here, the number on the top half of the circle indicates the item ID, and the number on the bottom half indicates the item category ID.
 (b) We randomly select two sessions (i.e., $\mathrm{Session}_A$ and $\mathrm{Session}_B$) for each user of the Taobao dataset and then merge them into one (i.e., $\mathrm{Merged\ Session}$). We use two metrics  (i.e., entropy and Gini index) \cite{zheng2021collaborative} to quantitatively evaluate the category diversification of these sessions, where a high entropy or low Gini index implies the great diversity.  }\label{fig1}
    \vspace{-1.3em}
\end{figure*}

However, existing work has overlooked some non-trivial issues when modeling different time-scale user interests. Firstly, most of these studies capture user interests via ubiquitous implicit feedback, which is susceptible to various factors, such as position bias \cite{jagerman2019model} and caption bias \cite{hofmann2012caption}, leading to \textit{unintentional behavioral noise}. This kind of noise is particularly harmful to the extraction of short-term user interests, which are usually extracted from a small set of most recent behaviors, thus making short-term interest modeling more vulnerable to unintentional behavioral noise. Secondly, these studies mostly capture long-term user interests via historical behaviors spanning several sessions, where a session is a sequence of behaviors in close temporal proximity  \cite{fang2020deep}. As shown in Figs.~\ref{fig1}(a) and~\ref{fig1}(b), we have noticed that \textit{user behaviors are highly homogeneous within a session and heterogeneous across sessions}, which implies that behaviors occurring in different sessions may be irrelevant to each other. Conventional long-term interest modeling usually ignores the multi-session structure of the historical sequence, so even if a certain behavior is not unintentional behavioral noise, it may introduce another form of noise by executing numerous ineffective feature interactions with behaviors in other sessions, ultimately impairing the mining of long-term user interests. We refer to this kind of noise as \textit{inter-session behavioral noise}.

There are several studies that have explored the denoising of SR models to alleviate the harmful impact of noise. Some early attention-based work \cite{luo2020collaborative,qin2021world} has utilized item-level relevance to the target item, down-weighting or directly removing items with low relevance. These methods solely rely on the target item as the denoising signals, although they filter out unintentional behaviors, they also unavoidably discard many low-relevance but non-noisy behaviors, resulting in the loss of potentially useful information. Another line of work \cite{lin2023self,zhou2022filter,du2023frequency} utilizes the sequence context as more reliable denoising signals to filter out noisy components that are inconsistent with the sequence semantics. However, these methods require rich within-sequence information and are less suitable for the denoising of short-term interest modeling, which relies on a few recent behaviors with limited information. Besides, they cannot deal with inter-session behavioral noise that is introduced by conventional long-term interest modeling.

To alleviate the negative effects of different time-scale noise when modeling different time-scale user interests, we propose a Long- and Short-term Interest Denoising Network (LSIDN), which employs different encoders and tailored denoising strategies to extract long- and short-term user interests, respectively, achieving both comprehensive and robust user modeling. Firstly, we extract long-term user interests from historical behaviors spanning several sessions. Due to the multi-session structure of the history sequence and the heterogeneity of inter-session behaviors, we extract intra-session interests separately and then capture the evolution of inter-session interests, thus avoiding the introduction of inter-session behavioral noise into long-term interest modeling. Secondly, we model the short-term user interests from the most recent behaviors. Due to the limited length of the current session and the homogeneity of intra-session behaviors, we resort to contrastive learning \cite{chen2020simple} with a novel data augmentation tailored for short sequences, thus alleviating the impact of unintentional behavioral noise on short-term interest modeling. Our homogeneous exchanging augmentation builds more informative and reliable positive pairs by introducing future data based on intra-session semantic consistency, mitigating data sparsity and information loss caused by traditional sequence-based augmentations \cite{yu2023self}. Note that the future data are the suffixed sub-sessions of the original current session and can only be accessed during training. Finally, we combine the long- and short-term interests in an adaptive way \cite{yu2019adaptive} for joint prediction. Our contributions are summarized as follows.
\vspace{-1.1mm}
\begin{itemize}
\setlength{\itemsep}{1.5pt}
\setlength{\parsep}{1.5pt}
\setlength{\parskip}{1.5pt}
\item[$ \bullet$ ] We are the first to indicate the negative effects of different time-scale noise in the sequential recommendation and propose a Long- and Short-term Interest Denoising Network, which employs distinct encoders and tailored denoising strategies to extract different time-scale user interests, thus achieving both comprehensive and robust user modeling.
 \item[$ \bullet$ ] We design a session-level interest extraction and evolution strategy to avoid introducing inter-session behavioral noise into the long-term interest modeling, and we also resort to contrastive learning equipped with a novel augmentation method to alleviate the impact of unintentional behavioral noise on short-term interest modeling.
\item[$ \bullet$ ] We conduct extensive experiments on two real-world datasets to show that LSIDN consistently outperforms various state-of-the-art models, and release the code at \url{https://github.com/zxyllq/LSIDN.git} for reproducibility.
\end{itemize}

\section{Related Work}
 
\subsection{Modeling Long- and Short-term Interests} 
Early deep neural network-based recommendation methods \cite{he2017neural,zhou2018deep,tang2018personalized,hidasi2015session} capture a holistic user interest without distinguishing between multiple interests. In realistic scenarios, the user's interests can be viewed over different time scales, including long- and short-term interests, and their combination will facilitate comprehensive user modeling \cite{ying2018sequential,lv2019sdm,yu2019adaptive,zheng2022disentangling,tan2021dynamic}. For example, SHAN \cite{ying2018sequential} employs a novel two-layer hierarchical attention network to learn the user's general taste and recent demand, respectively. SLi-Rec \cite{yu2019adaptive} improves the classical LSTM and proposes an attention-based fusion method to adaptively combine long- and short-term interests. CLSR \cite{zheng2022disentangling} adopts contrastive learning to disentangle long- and short-term user interests. However, these methods ignore the negative influence of noise when modeling different time scales of user interests.

\subsection{Denoising for Sequential Recommendation} 
Previous attention-based work \cite{luo2020collaborative,qin2021world} has utilized item-level relevance to the target item, down-weighting or directly removing items with low relevance. However, these methods unavoidably discard many low-relevance but non-noisy behaviors, resulting in the loss of potentially useful information. Recently, another line of work \cite{lin2023self,zhou2022filter,du2023frequency} has made further progress by utilizing the sequence context as more reliable denoising signals to filter out noise. For example, STEAM \cite{lin2023self} designs an item-wise corrector to explicitly correct the raw sequence. 
FMLP-Rec \cite{zhou2022filter} introduces a filter-enhanced MLP to filter out noise components from the item representation. FEARec \cite{du2023frequency} captures behavior patterns of different frequency bands and implicitly removes noise in the frequency domain. However, these methods fail to alleviate the negative effects of different time-scale noise in the sequential recommendation.

\subsection{Contrastive Learning} Contrastive learning \cite{yu2023self, yu2022improving} is a branch of self-supervised learning, and it adopts data augmentation to learn high-quality and generalizable representations. For example, $\mathrm{S^3}$-Rec \cite{zhou2020s3} uses random masks of attributes and sequences to maximize the mutual information over positive pairs. CL4SRec \cite{xie2022contrastive} and CLUE \cite{cheng2021learning} perform \textit{crop}, \textit{mask}, and \textit{reorder} on a sequence to obtain multiple views of the same behavior sequence. However, these traditional sequence-based augmentations may break essential item relationships or lead to fewer items, yielding less confident positive pairs, especially for short sequences.

 \section{Methodology}
 
\subsection{Problem Formulation}We denote user and item sets as $\mathcal{U}$ and $\mathcal{V}$, respectively. For each user ${u}\in\mathcal{U}$, we have his/her temporally ordered behavior sequence ${\boldsymbol{s}} =[{v}_1,{v}_2,\ldots,{v}_t,\ldots, {v}_{|\boldsymbol{s}|}]$, where $ {v}_t \in \boldsymbol{s}$ is the $t$-th interacted item of user $ {u}$. Given a user's history sequence with $T-1$ time steps, our model aims to predict the next-item ${v}_{T}$ that the user will interact with. 

\subsection{Overview} 
\begin{figure*}[htpb]
 \centering
\includegraphics[width=0.9\textwidth]{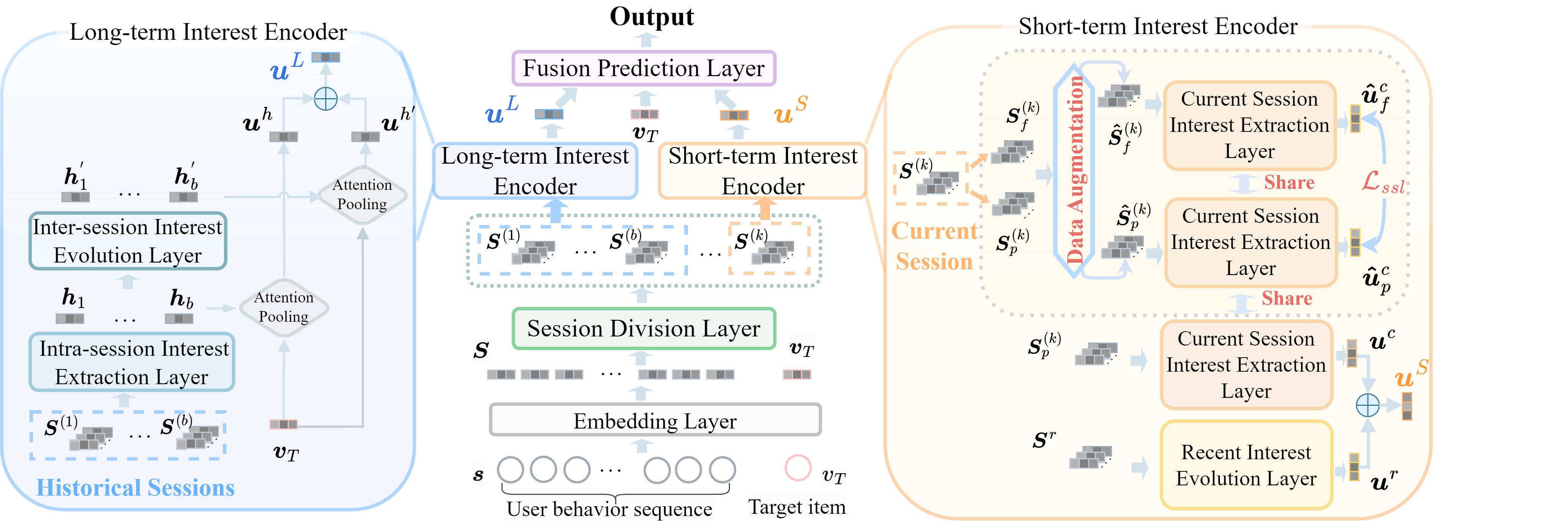}
 
\caption{ The architecture of our proposed model LSIDN. }\label{fig2_sdm}
 \vspace{-1.5em}
\end{figure*}

The architecture of our LSIDN is shown in Fig. \ref{fig2_sdm}. Firstly, we divide the user behavior sequence into sessions through the session division layer considering the multi-session structure of the historical sequence. Secondly, we use long-term and short-term interest encoders to model different time-scale user interests, respectively. Finally, we aggregate the different time-scale user interests through the fusion prediction layer and then calculate the interaction probability of the target item $v_T$. Key components and technical details are elaborated below.

\subsection{Session Division Layer}
User historical behaviors are not uniformly distributed over timestamps, and behaviors that are in close temporal proximity tend to be closely related,  so we divide the historical sequence into multiple sessions. Note that, unlike session-based recommendation models~\cite{fang2020deep} that focus on anonymous and individual sessions, our model is for the sequential recommendation that focuses on non-anonymous and cross-session sequences, where sessions are merely intermediate products of our model. 

Firstly, we create an item embedding matrix $\boldsymbol{E} \in \mathbb{R}^{|\mathcal{V}| \times d}$, where $d$  is the embedding size, and then $v_t$ and $\boldsymbol{s}$ are embedded as $\boldsymbol{v}_t \in \mathbb{R}^{1\times d}$ and  $\boldsymbol{S}\in\mathbb{R}^{|\boldsymbol{s}|\times d}$. Secondly, we adopt the session division layer, denoted as  $\mathrm{SessDiv}(\cdot)$, to divide the sequence into sessions. Specifically, given a division threshold $\omega$, for $\forall \boldsymbol{v}_t \in \boldsymbol{S}$, if the time interval $\Delta t$ between $\boldsymbol{v}_{t-1}$ and $\boldsymbol{v}_{t}$ is less than $\omega$, then $\boldsymbol{v}_{t-1}$ and $\boldsymbol{v}_t$ belong to the same session, otherwise, $\boldsymbol{v}_{t-1}$ and $\boldsymbol{v}_t$ belong to two successive sessions, respectively. As a result, $\boldsymbol{S}$ is divided as follows.
 \begin{equation}
 [\boldsymbol{S}^{(1)}, \boldsymbol{S}^{(2)},\ldots,\boldsymbol{S}^{(n)},\ldots, \boldsymbol{S}^{(k)}] =\mathrm{SessDiv}(\boldsymbol{S},\omega),
\end{equation}

\noindent where $k$ is the number of sessions contained in sequence $\boldsymbol{s}$,  $\boldsymbol{S}^{(n)}\in \mathbb{R}^{l\times d}$ is the $n$-th session embedding matrix,  and $l$ denotes the maximum session length. 

\subsection{Long-term Interest Encoder} After dividing the sequence into $k$ ($k$ is a dynamic value) sessions, we feed the earlier $b$ ($b \leq k$) sessions into the long-term interest encoder to allow instances to be trained in mini-batches. Then we elaborate on the technical details of the long-term interest encoder in the following.
 
\noindent \textbf{Intra-session Interest Extractor Layer.} We adopt a Transformer-based encoder \cite{vaswani2017attention} to extract the intra-session interest and denote it as $\mathrm{SessEnc}(\cdot)$, which is shared across $b$ sessions, i.e., $[\boldsymbol{S}^{(1)}, \boldsymbol{S}^{(2)},..., \boldsymbol{S}^{(b)}]$. Given the \textit{n}-th historical session $\boldsymbol{S}^{(n)}$, where $n \in [1,2,\ldots, b]$, we obtain the intra-session interest $\boldsymbol{h}_n$ as follows.
 \begin{equation}\label{SessEnc} 
    \boldsymbol{h}_n=\mathrm{AvgPooling}(\mathrm{SessEnc}(\boldsymbol{S}^{(n)})).
\end{equation}

 \noindent Then we can get $b$ intra-session interests $[\boldsymbol{h}_{1} , \boldsymbol{h}_{2}, \ldots, \boldsymbol{h}_{b}]$. By extracting session interests separately, we bypass the computation of self-attention over the whole sequence. It can be viewed as a form of local self-attention \cite{beltagy2020longformer} that helps to avoid inter-session behavioral noise, i.e., ineffective feature interactions between irrelevant behaviors occurring in different sessions, and thus improves both the efficiency and effectiveness of interest modeling.

\noindent \textbf{Inter-session Interest Evolution Layer.} We adopt GRU \cite{dey2017gate}, a variant of recurrent neural networks (RNNs), to capture the evolution of historical interest. Although RNNs perform well at capturing temporal dynamics, they suffer from inefficient inference time due to their low-parallelism architectures and tend to forget essential information when processing long sequences. 

Fortunately, we model the temporal dynamics between sessions rather than between individual items, the total number of sessions is much smaller than that of items, thereby significantly reducing inference time and relieving the information forgetting. Besides, a coarse-grained session representation compresses several fine-grained item representations, so that taking the session-level representations as input is equivalent to enlarging the receptive field of the RNN cell. We formulate inter-session interests as the hidden states of GRU as follows. 
 \begin{equation} \label{eq3}
[\boldsymbol{h}_{1}^{\prime}, \boldsymbol{h}_{2}^{\prime}, \ldots, \boldsymbol{h}_{b}^{\prime}] =\mathrm{GRU}([\boldsymbol{h}_{1} , \boldsymbol{h}_{2}, \ldots, \boldsymbol{h}_{b}]).
\end{equation}

\noindent Next, we adopt an \textbf{attention pooling} operation to conduct a soft alignment between the user's multiple session-level interests and the target item $\boldsymbol{v}^T$. For user's intra-session interests, i.e., $[\boldsymbol{h}_{1} , \boldsymbol{h}_{2}, \ldots, \boldsymbol{h}_{b}]$, we have
 \begin{small}
 \begin{align}\label{3.4}  
    & \boldsymbol{u}^{h}=\mathrm{AttnPool}(\boldsymbol{v}_T,[\boldsymbol{h}_{1} , \boldsymbol{h}_{2}, \ldots, \boldsymbol{h}_{b}]) =\sum_{i=1}^b a_i\boldsymbol{h}_i,\\ 
    \nonumber
    & a_i=\frac{exp(\boldsymbol{h}_i\boldsymbol{W}\boldsymbol{v}_T^{\top})}{\sum_{j=1}^b exp(\boldsymbol{h}_j \boldsymbol{W}\boldsymbol{v}_T^{\top})}, 
\end{align}
 \end{small}

\noindent where  $\boldsymbol{W}\in \mathbb{R}^{d \times d}$ are learnable parameters. Similarly, we also perform the above attention pooling operation on the user's inter-session interests, i.e., $[\boldsymbol{h}_{1}^{\prime}, \boldsymbol{h}_{2}^{\prime}, \ldots, \boldsymbol{h}_{b}^{\prime}]$, to obtain the aggregated representation $\boldsymbol{u}^{h^{\prime}}$. Finally, the user's long-term interest can be expressed as
\begin{equation} 
    \boldsymbol{u}^L=\mathrm{Concat}(\boldsymbol{u}^{h},\boldsymbol{u}^{h^{\prime}}).
\end{equation}

\subsection{Short-term Interest Encoder}

We use the current session  $\boldsymbol{S}^{(k)}$ for short-term interest modeling. However, some current sessions contain only one or two items and thus lack clues for user modeling. Hence, we take the latest $r\ (r\textless T-1)$ items, i.e., the recent sub-sequence $\boldsymbol{S}^r$, as the complementary input. The technical details of the short-term interest encoder are described below.

\noindent \textbf{Current Session Interest Extractor Layer. }We utilize an encoder that has the same structure as $\mathrm{SessEnc}(\cdot)$ in Eq. \ref{SessEnc} and denote it as $\mathrm{SessEnc}^{\prime}(\cdot)$. Then we get the current session interest $\boldsymbol{u}^c$ as follows.

\begin{equation}
\boldsymbol{u}^c=\mathrm{AvgPooling}(\mathrm{SessEnc}^{\prime}(\boldsymbol{S}^{(k)})).
\end{equation}

\begin{figure*}[htbp]
 
\centering
\includegraphics[width=0.8\textwidth]{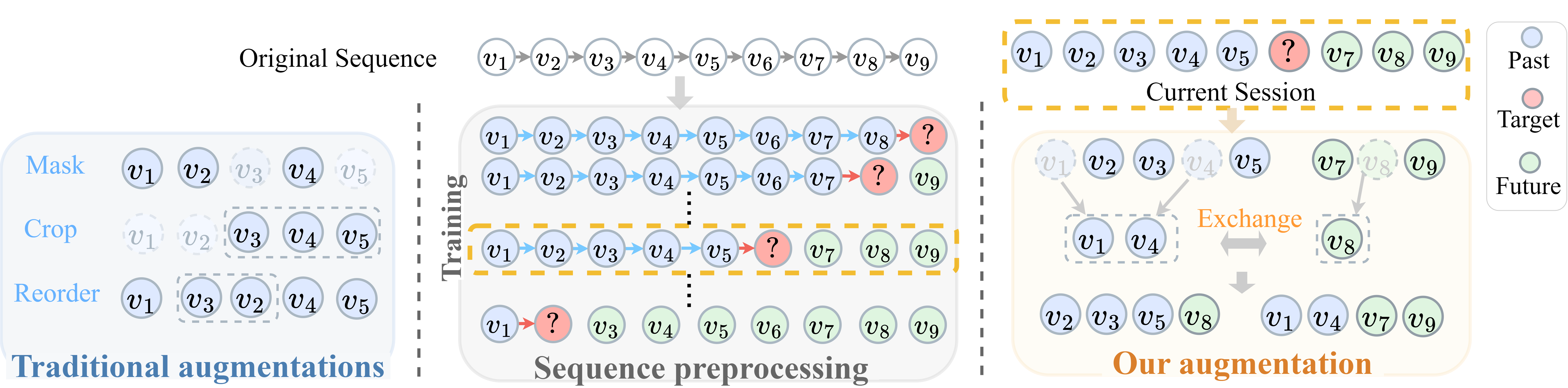}
 
\caption{ Illustration of our proposed augmentation method. }\label{fig3}
 \vspace{-1.5em}
\end{figure*}

\noindent However, the limited length makes the current session more sensitive to unintentional behavior noise, so we adopt contrastive learning to enhance the robustness of the representation. However, as shown in the left part of Fig. \ref{fig3}, traditional sequential augmentations \cite{yu2023self} may break essential item relationships or lead to fewer items in sequences, thus yielding less confident positive pairs and impairing the quality of the learned representation, especially for short sequences. Considering that
 short-term interest modeling usually relies on limited recent behaviors, we propose a new data augmentation method tailored for short sequences, which uses the suffixes of the original current session as additional information to create reliable self-supervised signals.

As shown in the center part of Fig. \ref{fig3}, most SR methods adopt a sequence preprocessing method \cite{tan2016improved} to enhance training by generating a collection of prefixes with corresponding new targets as new training instances, while discarding the suffixes, i.e., future data after the corresponding new targets. As we can see, the future data are accessible during training and worth exploiting. Therefore, in the training phase, we specify that the current session $\boldsymbol{S}^{(k)}$  contains a past sub-session $\boldsymbol{S}^{(k)}_p$ before the target item and a future sub-session $\boldsymbol{S}^{(k)}_f$ after it; while during inference, the future data are inaccessible, thus  $\boldsymbol{S}^{(k)}=\boldsymbol{S}^{(k)}_p$. Due to the homogeneity of intra-session behaviors, the user's interests in the past and future sub-sessions are highly similar, so we randomly exchange items from the two sub-sessions to build positive pairs. As shown in the right part of Fig. \ref{fig3}, firstly, we randomly select a certain proportion of items from the past and future sub-sessions, respectively. Secondly, we exchange these selected items and merge them with the reserved items of the other side. Finally, we sort the two merged sessions by timestamp to obtain the augmented positive pairs. This homogeneous exchanging augmentation method is formulated as
\begin{small}
\begin{align}
    \nonumber
    & \boldsymbol{S}^{(k)}_{p\prime} = \mathcal{T} _{\mathrm{select}}(\boldsymbol{S}^{(k)}_p,\gamma),\boldsymbol{S}^{(k)}_{f \prime} =\mathcal{T}_{\mathrm{select}}(\boldsymbol{S}^{(k)}_f,\gamma),\\
    \nonumber
     & \boldsymbol{S}^{(k)}_{p\prime \prime} =\mathcal{T} _{\mathrm{delete}}(\boldsymbol{S}^{(k)}_p,\boldsymbol{S}^{(k)}_{p\prime}),\boldsymbol{S}^{(k)}_{f \prime \prime} =\mathcal{T}_{\mathrm{delete}}(\boldsymbol{S}^{(k)}_f,\boldsymbol{S}^{(k)}_{f\prime}),\\
     & \boldsymbol{\hat{S}}^{(k)}_{p}=\mathcal{T} _{\mathrm{sort}}(\boldsymbol{S}^{(k)}_{f\prime} \oplus \boldsymbol{S}^{(k)}_{p\prime \prime}),\boldsymbol{\hat{S}}^{(k)}_{f}=\mathcal{T} _{\mathrm{sort}}(\boldsymbol{S}^{(k)}_{p\prime}\oplus\boldsymbol{S}^{(k)}_{f\prime \prime}),
\end{align}
\end{small}

\noindent where $\gamma$ denotes the selection ratio and $\oplus$ denotes the merge operation. During training, let $\mathcal{B}$ denotes a mini-batch of training instances, which includes $|\mathcal{B}|$ current sessions, each session containing a past and a corresponding future sub-session. Given the \textit{i}-th current session, both its past and future sub-sessions, i.e., $S^{(k)}_{p_i}$ and  $S^{(k)}_{f_i}$, will adopt homogeneous exchanging augmentation method to obtain positive pairs, i.e., $\hat{S}^{(k)}_{p_i}$ and  $\hat{S}^{(k)}_{f_i}$, which are then fed into $\mathrm{SessEnc}^{\prime}(\cdot)$ to get the representations, i.e., $\hat{\boldsymbol{u}}_{p_i}^c$ and $\hat{\boldsymbol{u}}_{f_i}^c$. Following \cite{chen2020simple}, $(\hat{\boldsymbol{u}}_{p_i}^c,\hat{\boldsymbol{u}}_{f_i}^c)$ is treated as a positive pair, while $\hat{\boldsymbol{u}}_{p_i}^c$ with the other $2(|\mathcal{B}|-1)$ augmented representations are considered as negative pairs for this pair. We adopt the NT-Xent loss \cite{chen2020simple} for optimization as follows.
\begin{small}
 \begin{equation}\label{3.12}      \mathcal{L}_{ssl}(\boldsymbol{\hat{u}}_{p_i}^{c},\boldsymbol{\hat{u}}_{f_i}^{c} )
      = -\mathrm{log} \frac{\mathrm{exp}(\mathrm{sim}( \boldsymbol{\hat{u}}_{p_i}^{c},\boldsymbol{\hat{u}}_{f_i}^{c} )/\tau)}{\sum_{j=1}^{|\mathcal{B}|} \mathbb{I}_{[j \neq  i]} \mathrm{exp}(\mathrm{sim}( \boldsymbol{\hat{u}}_{p_i}^{c},\boldsymbol{\hat{u}}_{*_j}^{c})/ \tau)},
\end{equation}
\end{small}

\noindent where $\mathbb{I}_{[j \neq i]}\in \{0,1\}$ is an indicator function, $\mathrm{sim(\cdot)}$ is a dot product, * denotes $p$ or $f$, and $\tau$ is a temperature index. It is worth mentioning that we only calculate $\mathcal{L}_{ssl}$ for training instances that contain both past and future sub-sessions.

\noindent \textbf{Recent Interest Evolution Layer.} We take the latest $r\ (r\textless {T-1})$ items as the recent sub-sequence, i.e., $\boldsymbol{S}^r=[\boldsymbol{v}_{T-r-1},\boldsymbol{v}_{T-r},...,\boldsymbol{v}_{T-1}]$. We employ Time4LSTM \cite{yu2019adaptive} combined with the attention pooling in Eq. \ref{3.4} to obtain recent user interest $\boldsymbol{u}^r$ as follows. 
 \begin{equation}
 \boldsymbol{u}^r= \mathrm{AttnPool}( \boldsymbol{v}_T,\mathrm{Time4LSTM}(\boldsymbol{S}^r) ).
 \end{equation}
We set the recent interest evolution layer as an option to be used only when the current session is extremely short, and also set $r$ as small as possible to balance effectiveness and efficiency. Finally, the short-term user interest is expressed as: 
    \begin{equation}\label{3.13}
    \boldsymbol{u}^S= \mathrm{Concat}(\boldsymbol{u}^{r}, \boldsymbol{u}^{c}).
\end{equation}
\subsection{Fusion Prediction Layer}  
Instead of using simple aggregators (e.g., summing and concatenation), we \textit{adaptively} aggregate long- and short-term interests for information fusion, considering the target item $\boldsymbol{v}_{T}$ and the trade-off between different time-scale interests  \cite{yu2019adaptive}.
\begin{small}
    \begin{align}
    \nonumber
    & \alpha =  \mathrm{Sigmoid}(\boldsymbol{W}^m\mathrm{Concat}(\boldsymbol{u}^L,\boldsymbol{u}^S,\boldsymbol{v}_{T})+\boldsymbol{b}_m),\\& \boldsymbol{u}^{LS}=\alpha \boldsymbol{u}^L+(1-\alpha)\boldsymbol{u}^S,
\end{align}
\end{small}

 \noindent where $\boldsymbol{W}^m$ and $\boldsymbol{b}_m$ are learnable parameters. Finally, we use the widely adopted two-layer MLP \cite{zheng2022disentangling} to predict the score of the target item $\boldsymbol{v}_T$.
 
\begin{equation}\label{3.16}
    \hat{y}_{u,v}=\mathrm{Sigmoid}(\mathrm{MLP}(\mathrm{Concat}(\boldsymbol{u}^{LS},\boldsymbol{v}_T))).
\end{equation}

\noindent We use the negative log-likelihood loss \cite{yu2019adaptive,zheng2022disentangling}  as follows. 
 \begin{small}
\begin{equation}
    \mathcal{L}_{main}(u,v)=-\frac{1}{N}\sum\nolimits_{v \in \mathcal{O}} y_{u,v}\mathrm{log}(\hat{y}_{u,v})+(1-y_{u,v})\mathrm{log}(1-\hat{y}_{u,v}),
\end{equation}
 \end{small}
 
\noindent where $\mathcal{O}$ is the set of training instances, containing a positive item $v_T$ that the user has interacted with, and $N-1$ negative items that are sampled within a batch. Finally, we use a multi-task strategy to optimize the above two objectives:
\begin{equation}
    \mathcal{L}=\mathcal{L}_{main}+\lambda \mathcal{L}_{ssl}+ \beta||\Theta||_2,
\end{equation}
 where $\Theta$ denotes the set of trainable parameters, $||\cdot||_2$ denotes the $L2$ regularization, $\lambda$ and $\beta$ are used to control the strength of  $\mathcal{L}_{ssl}$ and $L2$ regularization.

\section{Experiments}
 
\subsection{Experimental Settings} We elaborate on the details of experimental setups as follows.

\noindent \textbf{Datasets.} We conduct experiments on two public e-commerce datasets, i.e., \textbf{Taobao}\footnote{\url{https://tianchi.aliyun.com/dataset/649}} and \textbf{Fliggy}\footnote{\url{https://tianchi.aliyun.com/dataset/113649}}. Both of them include behaviors such as click, favorite, add-to-cart, and purchase. Details of the datasets are listed in Table \ref{tab1}. The Taobao dataset, collected from the largest online shopping platform in China, includes user-item interactions from 25/11/2017 to 4/12/2017. We use interactions before 1st Dec for training and interactions from the last 2 days for validation and testing, respectively. The Fliggy dataset, provided by one of the most popular online travel platforms in China, contains user-item interactions from 3/6/2019 to 3/6/2021. We use the first 18 months of interactions for training, the following 3 months for validation, and the final 3 months for testing. We leverage the same preprocessing and negative sampling methods as \cite{zheng2022disentangling}.

\noindent \textbf{Baselines and Metrics. }We compare LSIDN with recommendation models of various research lines: (\rmnum{1}) \textbf{simple holistic interest models} that neither distinguish between different time-scale user interests nor consider denoising, i.e., NCF \cite{he2017neural}, DIN \cite{zhou2018deep}, Caser \cite{tang2018personalized}, and GRU4Rec \cite{hidasi2015session}; (\rmnum{2}) \textbf{long- and short-term interest models} without denoising components, i.e.,  SLi-Rec \cite{yu2019adaptive} and CLSR \cite{zheng2022disentangling}; (\rmnum{3}) \textbf{denoising sequential models} that do not distinguish between different time-scale user interests, i.e., FMLP-Rec \cite{zhou2022filter} and FEARec \cite{du2023frequency}. We evaluate these models with two accuracy metrics, i.e., AUC and GAUC, and two ranking metrics, i.e., MRR and NDCG@K, here $K=\{5,10\}$.
 
\noindent \textbf{Implementation Details.} For a fair comparison, we implement all models with the Microsoft Recommenders framework \cite{argyriou2020microsoft} that is based on TensorFlow. We initialize all models with a normal distribution in the range $[-0.01,0.01]$ and use the early stopping strategy of \cite{zheng2022disentangling}. We set the embedding size, the batch size, and the maximum sequence length to 40, 500, and 50 for all models. Particularly for LSIDN, we set $b=5$ and $l=10$ for both datasets. We use a grid search to find the best hyper-parameters. Finally, $\beta$, $\lambda$, $\gamma$, and $r$ are respectively set to 0.1, 0.1, 0.4, and 30 for Taobao and Fliggy; $\tau$ is 0.2 for Taobao and 0.8 for Fliggy, $\omega$ is 360 minutes for Taobao and 5 days for Fliggy.

\subsection{Performance Comparison}
 \begin{table}[htpb]
 \setlength{\tabcolsep}{6.5pt} 
  \centering
   \resizebox{0.5\textwidth}{!}{ 
    \begin{tabular} {l c c c c}
     \toprule 
    Dataset &   \#Users & \#Items & \#Interactions &  Avg. length \\
     \midrule
    Taobao & 36,915 & 64,138 &   1,471,155 & 39.85 \\
    Fliggy & 180,232 & 50,355 & 5,781,867 & 22.58\\    
     \bottomrule
  \end{tabular}
 
  }
  \caption{Statistics of Datasets }  \label{tab1}
 \vspace{-1.8em}
\end{table} 
We list the performance of all models in Table \ref{tab2} and obtain the following observation. \textbf{Firstly}, the long- and short-term interest models, i.e., SLi-Rec and CLSR, outperform the simple holistic interest models in most cases, demonstrating the superiority of jointly modeling different time scales of user interests. \textbf{Secondly}, the denoising sequential models outperform the above non-denoising models in most cases, indicating the importance of denoising components. \textbf{Thirdly}, the proposed LSIDN exceeds all baselines by a large margin in all cases. This is attributed to the combination of different time-scale user interests and the implementation of corresponding time-scale denoising strategies, allowing for more comprehensive and robust user modeling.
 \begin{table*}[h]

   \centering
   \setlength{\tabcolsep}{1.8pt}
     \resizebox{0.8\textwidth}{!}{
     \begin{tabular}{l|ccccc|ccccc} 
         \toprule
          & \multicolumn{5}{c|}{Taobao} & \multicolumn{5}{c}{Fliggy}\\
         Model & AUC  & GAUC & MRR & NDCG@5 &NDCG@10& AUC  & GAUC & MRR & NDCG@5 &NDCG@10 \\
         \midrule
    
         NCF&0.7159&0.7196&0.1405&0.1231&0.1625&0.6622&0.6837&0.115&0.0951&0.1311\\
         DIN&0.7572&0.8552&0.3250&0.3355&0.3840&0.7695&0.8434&0.2898&0.2945&0.3444 \\

         Caser&0.8538&0.8607&0.3944&0.4055&0.4440&0.8075&0.8516&0.2910&0.3033&0.3526\\
         GRU4Rec&0.8699&0.8652&0.4307&0.4419&0.4777&0.8448&0.8587&0.3269&0.3353&0.3794\\
          \midrule
           
     SLi-Rec&0.8843&0.8816&0.3999&0.4143&0.4552&0.8622&0.8709&0.3175&0.3237&0.3705\\
     CLSR&0.9049&0.9052&0.4652&0.4817&0.5182&0.8751&0.8858&0.3436&0.3512&0.3981\\
        \midrule
    FMLP-Rec&0.8910&0.9158&0.5529&0.5696&0.5957&0.8782&0.9176&0.5547&0.5714&0.6011\\
    FEARec& 0.9155& 0.9179 & 0.5551& 0.5763 & 0.6070& 0.9063 & 0.9187 & 0.5581 & 0.5748 & 0.6055\\
    \midrule
    LSIDN&\textbf{0.9218}&\textbf{0.9213}&\textbf{0.5752}&\textbf{0.5931}&\textbf{0.6211}&\textbf{0.9082}&\textbf{0.9206}&\textbf{0.5741}&\textbf{0.5878}&\textbf{0.6144}\\
       \bottomrule 
    \end{tabular}
    }
  \caption{Performance comparisons of all models on both datasets. The best result in each column is in \textbf{bold}.}\label{tab2}

\end{table*}

\subsection{Ablation Study}
To demonstrate how each key component of LSIDN impacts overall performance, we design several LSIDN variants by removing certain components for comparisons, including (1) \textbf{w/o LD}: removing long-term denoising strategy (i.e., directly feeding the user behavior sequence into the encoder without session division), (2) \textbf{w/o SD}: removing short-term denoising components (i.e., removing the contrastive learning), (3) \textbf{w/o RI}: without the recent interest evolution layer , (4) \textbf{w/o CI}:  without the current session interest extraction layer, (5) \textbf{w/o LI}: without the long-term interest encoder, (6) \textbf{w/o\textit{ }SI}: without the short-term interest encoder. 
   \begin{table*}[htbp]
 
   \setlength{\tabcolsep}{4pt}

    \centering
    \setlength{\tabcolsep}{11pt}
    \resizebox{0.78\textwidth}{!}{ 
    \begin{tabular} {l |ll |ll} 
         \toprule
       \multirow{2}{*}{Method}   & \multicolumn{2}{c|}{Taobao} & \multicolumn{2}{c}{Fliggy}\\
            & GAUC  &NDCG@10  & GAUC  &NDCG@10 \\
         \midrule
        LSIDN &\textbf{0.9213}&\textbf{0.6211}&\textbf{0.9206}&\textbf{0.6144}\\
        \hline
        (1) w/o LD &0.8987 (-2.45\%) &0.5796 (-6.68\%)&0.9153 (-0.58\%)&0.6022 (-1.99\%)\\
        (2) w/o SD &0.9155 (-0.63\%)&0.5967 (-3.93\%)&0.9189 (-0.19\%)&0.6139 (-0.08\%)\\
        (3) w/o RI &0.9185 (-0.30\%)&0.6185 (-0.42\%)&0.9201 (-0.05\%)&0.6135 (-0.15\%)\\
        (4) w/o CI &0.9125 (-0.96\%)&0.5857 (-5.70\%)&0.8697 (-5.53\%)&0.4513 (-26.55\%)\\
        
        (5) w/o LI &0.9086 (-1.38\%)&0.6150 (-0.98\%)&0.9072 (-1.46\%)&0.5998 (-2.38\%)\\
        (6) w/o SI &0.8619 (-6.45\%)&0.4588 (-26.13\%)&0.8317 (-9.66\%)&0.3752 (-38.93\%)\\
       \hline 
    \end{tabular}
    }
        \caption{Ablation study of LSIDN. Percentages in parentheses indicate reductions relative to the original LSIDN.  }\label{Ablation}
        \vspace{-1.2em}
    \end{table*}

 From Table \ref{Ablation}, we find that removing any key component of LSIDN will result in performance degradation. Besides, we have the following observations. \textbf{Firstly}, the short-term interest encoder contributes more to performance than the long-term interest encoder, meanwhile, for the two components of the short-term interest encoder, i.e., current session interest extraction layer and the recent interest evolution layer, the former is more important than the latter. These observations are more evident in Fliggy, which may be due to the low-frequency nature of the travel scenario, making user modeling more reliant on the most recent behaviors. \textbf{Secondly}, our proposed different time-scale denoising strategies are more effective in Taobao. This may be because, in the shopping scenario, the goods are more abundant and diverse, thus user behaviors are more susceptible to various distractions and also more heterogeneous across sessions, which exacerbates the noise problem and makes our denoising strategies more useful.

\subsection{Long-  and Short-term Interest Analysis}
Taking the Taobao dataset as an example, we conduct further experiments to thoroughly analyze whether LSIDN captures \textit{effective} and \textit{meaningful} long- and short-term interest representations, respectively, by comparing LSIDN with two other long- and short-term interest models, i.e., CLSR and SLi-Rec.

\begin{figure*}[htpb] 
 \centering
 \vspace{-1.5em}
    \begin{minipage}[t]{0.184 \linewidth}
        \centering
        \includegraphics[width=\linewidth]{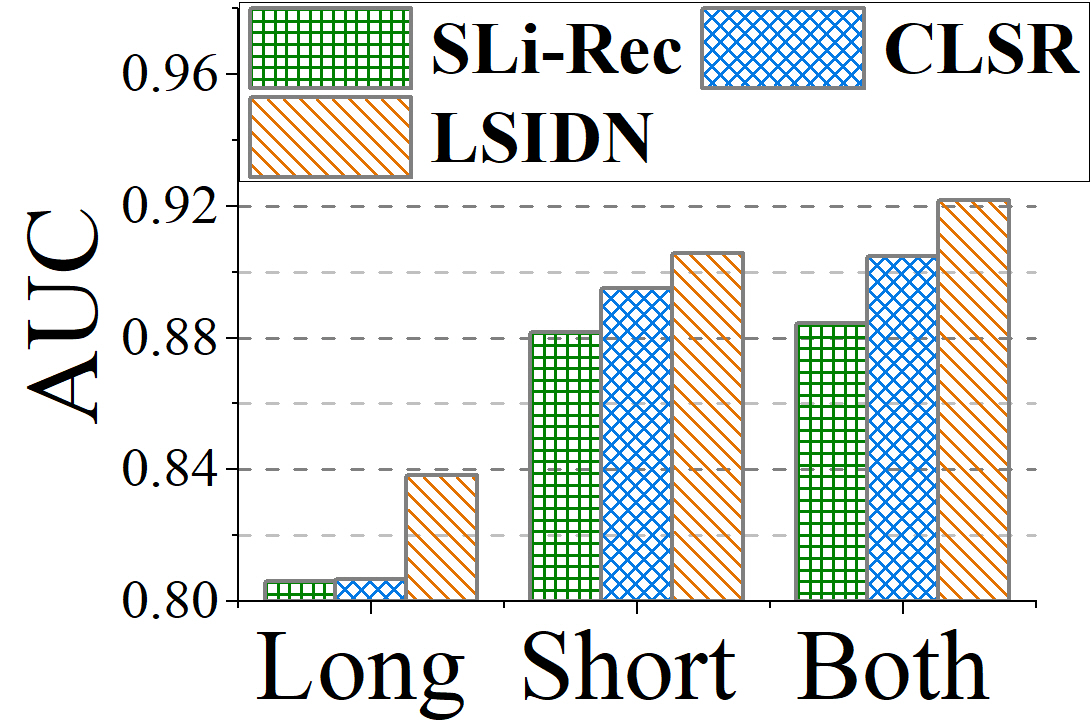}
        \centerline{(a)}
    \end{minipage} 
    \hspace{.0005in}%
    \begin{minipage}[t]{0.184 \linewidth}
        \centering
        \includegraphics[width=\linewidth]{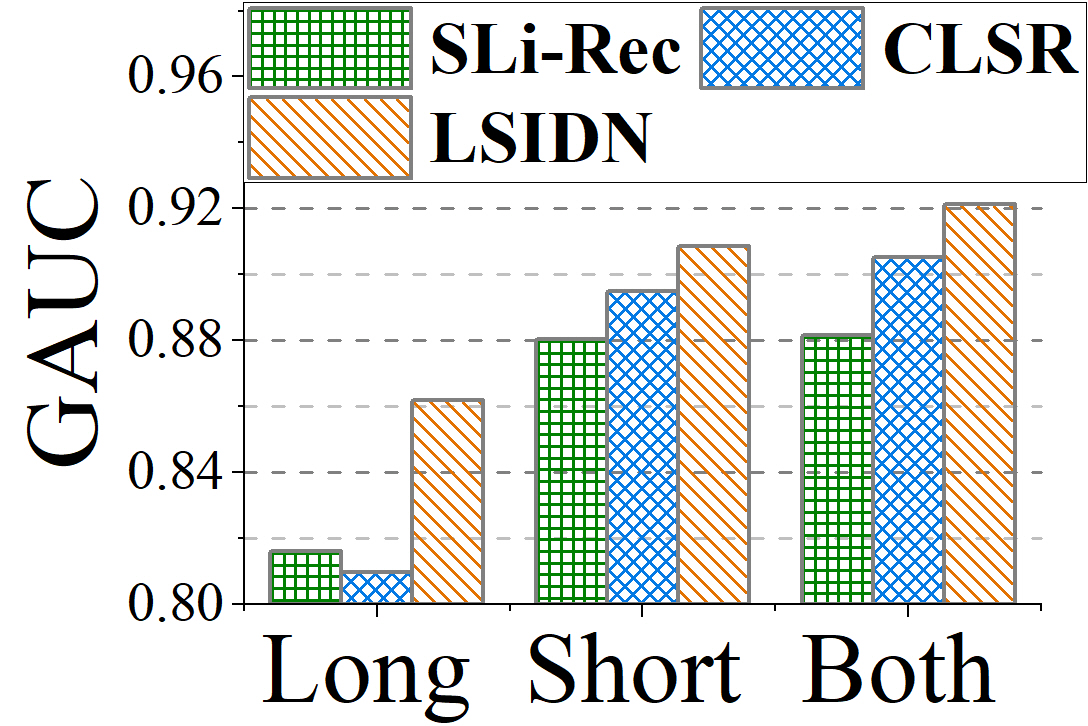}
        \centerline{(b)}
    \end{minipage}
    \begin{minipage}[t]{0.20\linewidth}
        \centering
        \includegraphics[width=\linewidth]{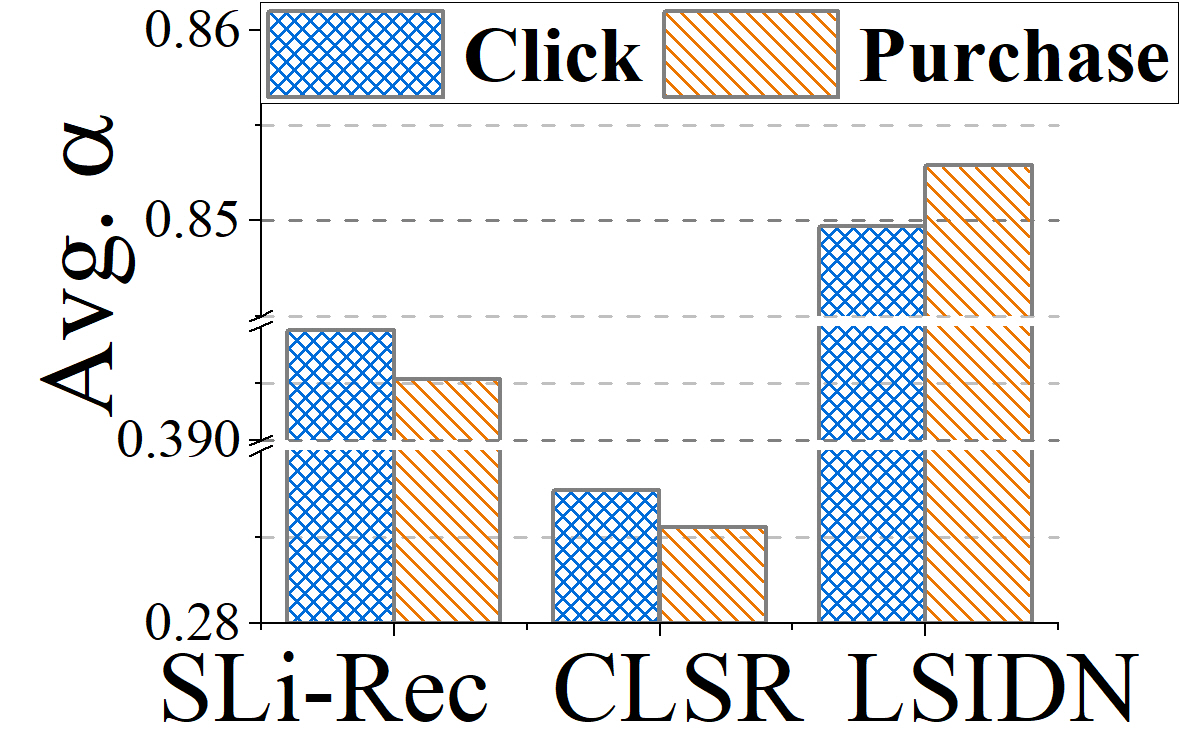}
        \centerline{(c)}
    \end{minipage}
    \begin{minipage}[t]{0.20\linewidth}
        \centering
        \includegraphics[width=\linewidth]{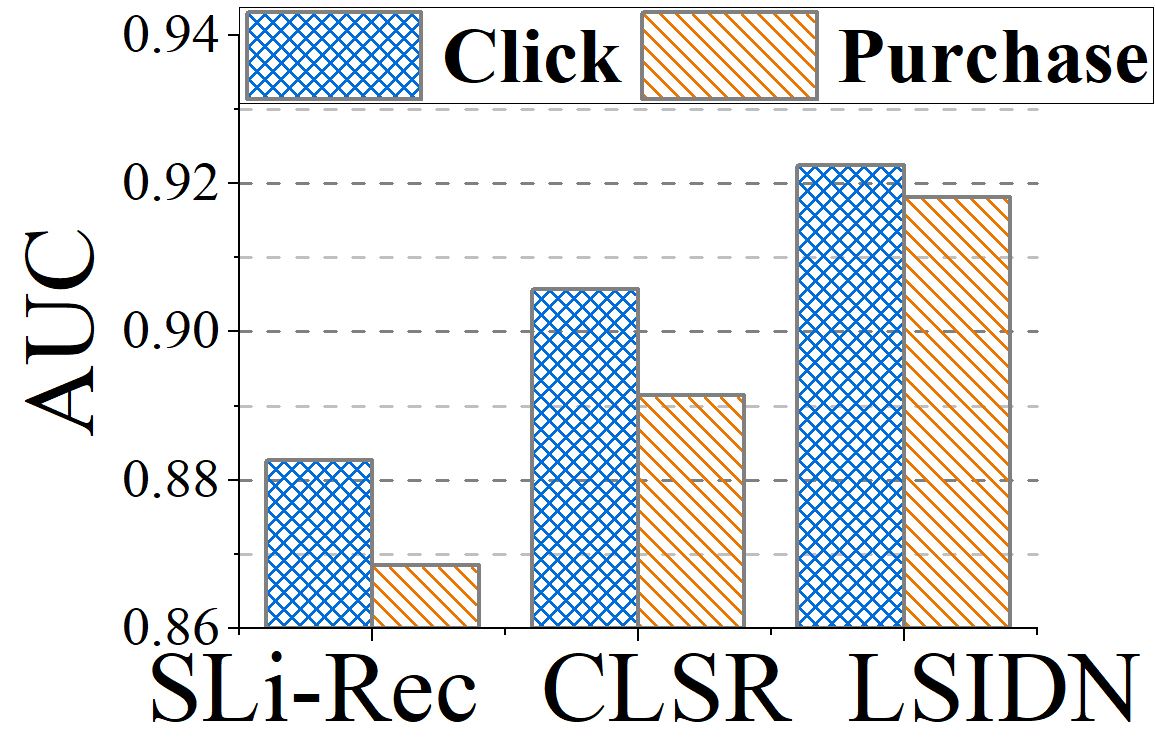}
        \centerline{(d)}
    \end{minipage}
    \begin{minipage}[t]{0.20\linewidth}
        \centering
        \includegraphics[width=\linewidth]{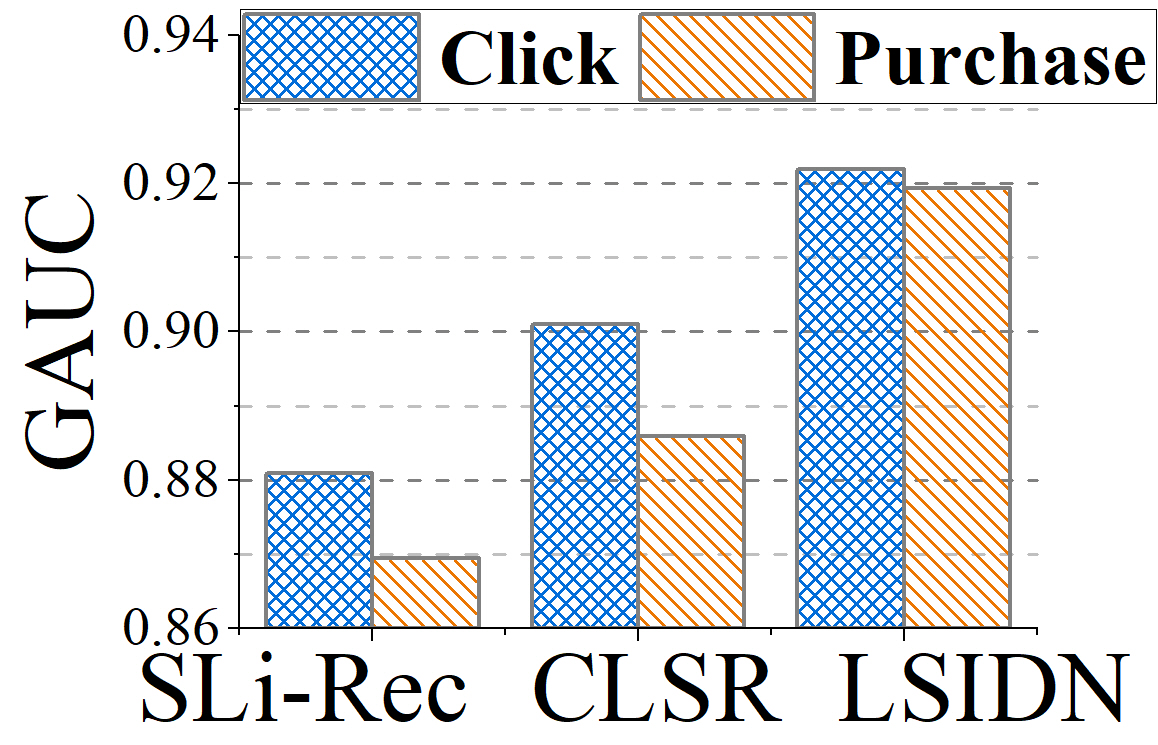}
        \centerline{(e)}
    \end{minipage}
 
    \caption{Performance comparison of long- and short-term interest models.}\label{fig4}
\end{figure*}

\textbf{Firstly}, given each of the three models, we separately construct two variants, i.e., \textbf{Long} (removing the short-term interest encoder) and \textbf{Short }(removing the long-term interest encoder), and we denote the original model as \textbf{Both}. From Figs. \ref{fig4}(a) and \ref{fig4}(b), we can see that LSIDN outperforms SLi-Rec and CLSR in all cases, revealing that LSIDN learns more \textit{effective} long- and short-term interests than the other two models. \textbf{Secondly}, we find that previous work \cite{grbovic2018real} claims that expensive behaviors (i.e., purchases) tend to be driven more by stable long-term user preferences, whereas cheap behaviors (i.e., clicks) are more likely to imply variable short-term user intentions; also, we find that SLi-Rec, CLSR, and LSIDN all have an adaptive weight $\alpha$, whose value depends on the sequence context and reflects the importance of the user's long-term interest. Inspired by this, we use each of the three previously trained models to make predictions for click and purchase items, respectively. We believe that if a certain one of the three models captures meaningful long- and short-term interests, then the Avg. $\alpha$ of the model on purchase items should be larger than its Avg. $\alpha$ on click items. From Fig. \ref{fig4}(c), we find that only LSIDN meets our expectations, suggesting that LSIDN has learned more \textit{meaningful} long- and short-term interests than SLi-Rec and CLSR. And from Figs. \ref{fig4}(d) and \ref{fig4}(e), LSIDN outperforms the other two models in both the purchase and click datasets, further demonstrating its superiority in long- and short-term user modeling.

\subsection{Robustness Analysis}
To evaluate the robustness of LSIDN and other methods, we modify the Taobao datasets as follows: contaminate the training and validation sets by adding a proportion of adversarial instances (i.e., 10\%, 20\%, and 30\% negative user-item interactions), while keeping the test set unchanged.

 \begin{figure*}[htbp] 
 \centering
    \begin{minipage}[t]{0.28 \linewidth}
        \centering
        \includegraphics[width=\linewidth]{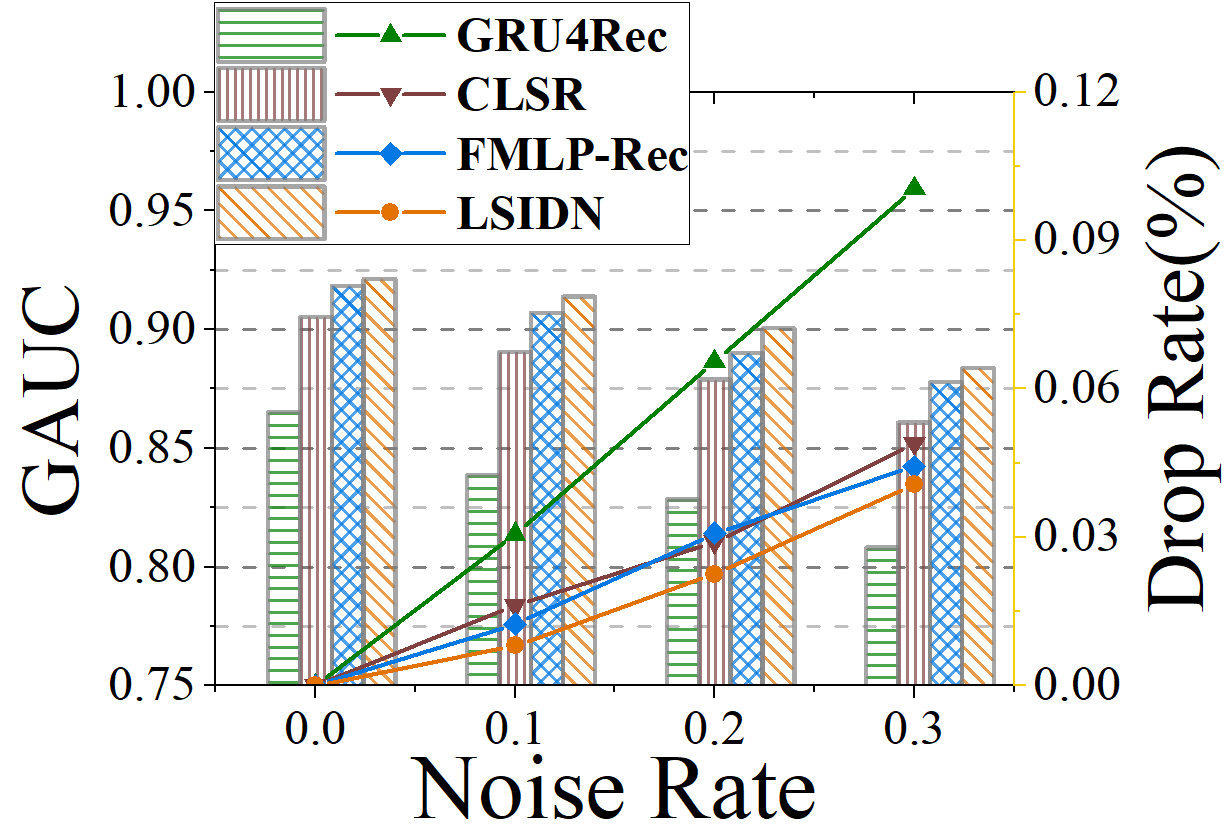 }
        \centerline{(a)}
    \end{minipage}%
    \hspace{.0005in}
    \begin{minipage}[t]{0.28 \linewidth}
        \centering
        \includegraphics[width=\linewidth]{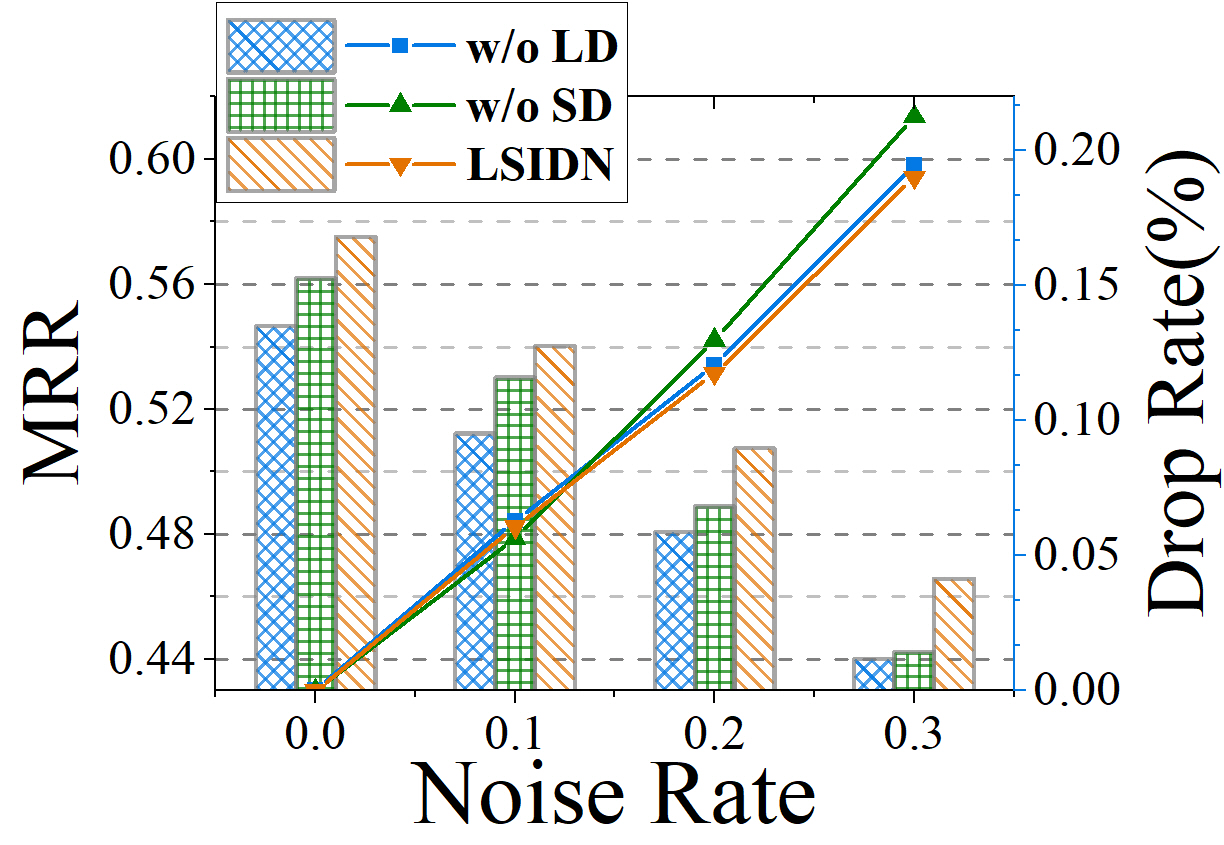}
        \centerline{(b)}
    \end{minipage}
    \begin{minipage}[t]{0.28\linewidth}
        \centering
        \includegraphics[width=\linewidth]{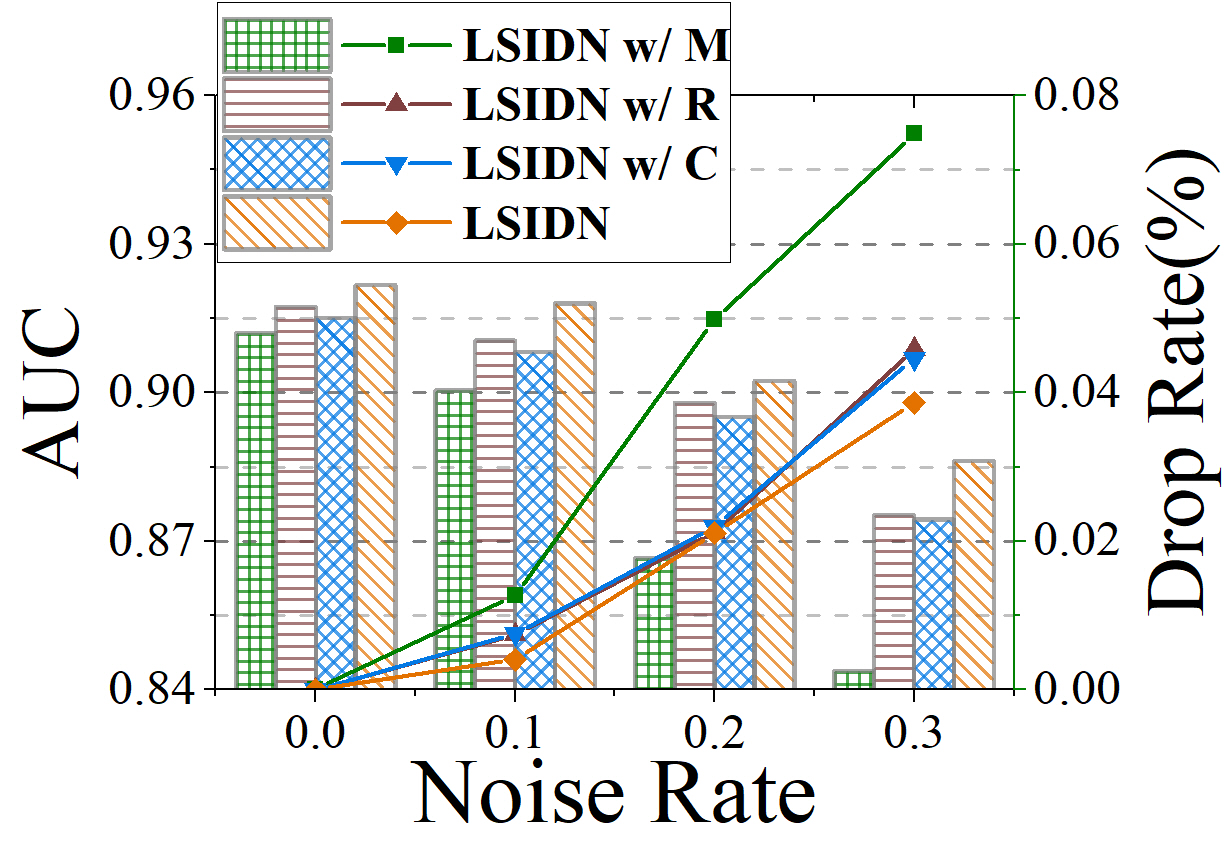}
        \centerline{(c)}
    \end{minipage}
 
    \caption{Robustness analysis of different methods. The bar represents the evaluation metrics, and the line represents the percentage of performance decrease, where a smaller drop rate indicates better noise resistance.
}\label{fig5}
 \vspace{-1.2em}
\end{figure*}
 
 \begin{figure*}[htbp]
   \vspace{-1.2em}
 \centering
    \begin{minipage}[t]{0.24 \linewidth}
        \centering
        \includegraphics[width=\linewidth]{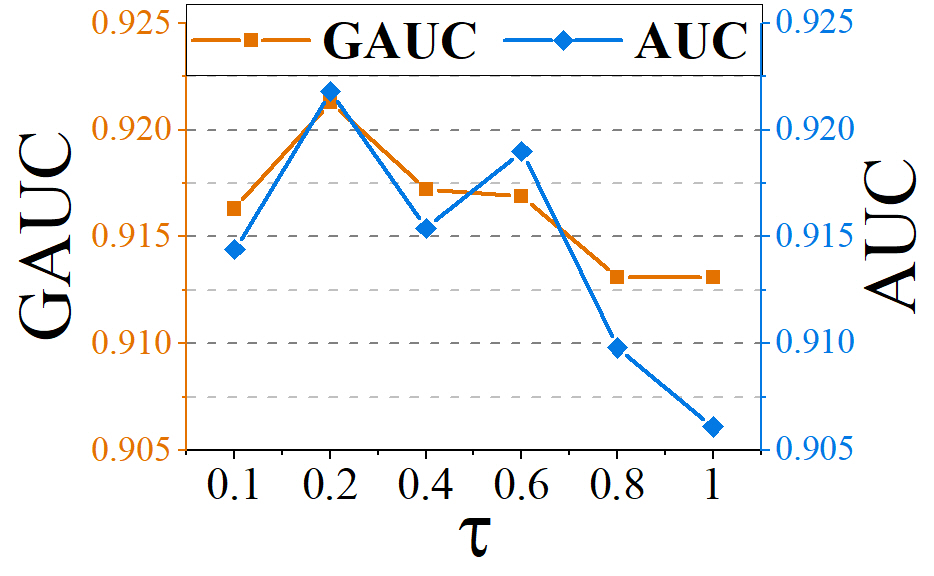 }
        \centerline{(a)}
    \end{minipage}
    \hspace{.004in}
    \begin{minipage}[t]{0.24 \linewidth}
        \centering
        \includegraphics[width=\linewidth]{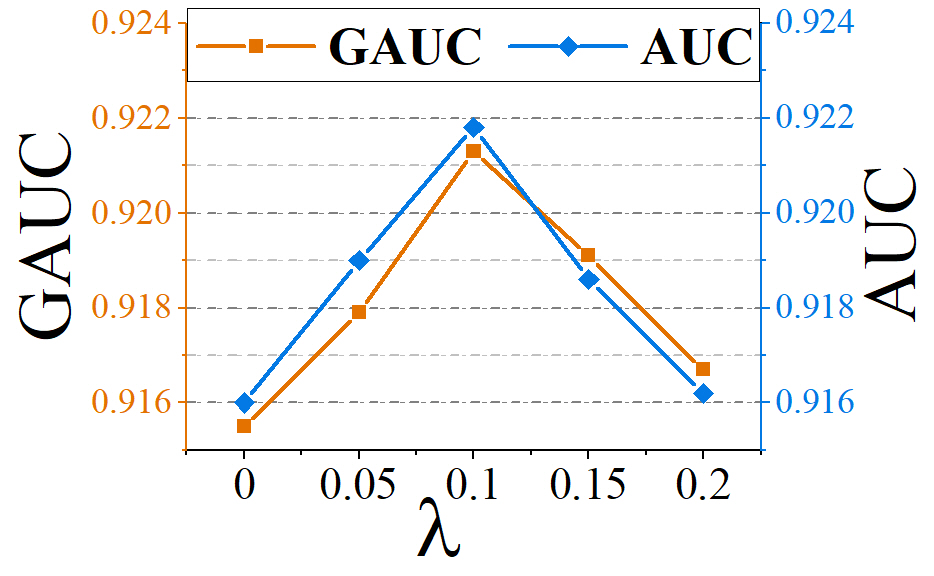}
        \centerline{(b)}
    \end{minipage}
      \hspace{.001in}
    \begin{minipage}[t]{0.24\linewidth}
        \centering
        \includegraphics[width=\linewidth]{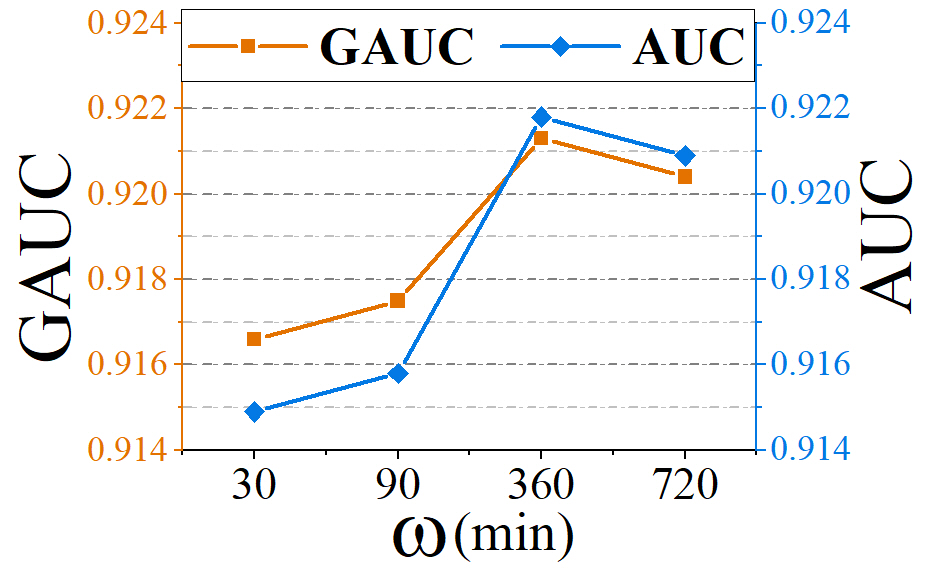}
        \centerline{(c)}
    \end{minipage}
 
    \caption{Hyper-parameter analyses of LSIDN on the Taobao dataset.
}\label{fig7}
  \vspace{-1.2em}
\end{figure*}

\textbf{Firstly}, we compare the robustness of LSIDN and several baseline models. As shown in Fig. \ref{fig5}(a), the simple holistic interest model, i.e., GRU4Rec, shows poor noise resistance, while FMLP-Rec and CLSR perform better, indicating that either denoising designs or joint modeling of different time-scale interests can improve model robustness. Importantly, LSIDN performs best in all cases due to its robust and comprehensive user modeling. \textbf{Secondly}, we investigate the effect of different time-scale denoising strategies employed in LSIDN. As shown in Fig. \ref{fig5}(b), we can see that after removing the denoising strategy in short-term interest modeling  (i.e., w/o SD), performance declines faster as the noise rate increases, proving the effectiveness of contrastive learning in denoising; after removing the denoising strategy in long-term interest modeling (i.e., w/o LD), the evaluation metric significantly decreases due to the introduced inter-session interference, but the noise resistance is comparable to original LSIDN due to reserved short-term denoising component. \textbf{Finally}, we replace our augmentation with traditional augmentation methods (i.e., crop, mask, and reorder) for comparisons, denoted as LSIDN w/ C, LSIDN w/ M, and LSIDN w/ R, respectively. As shown in Fig. \ref{fig5}(c), we can see that the original LSIDN outperforms its variants in terms of evaluation metrics and noise resistance, further demonstrating the superiority of our augmentation method.

\subsection{Hyper-parameter Sensitivity} 

We investigate the effect of three important hyperparameters on model performance, including (\rmnum 1) the temperature index $\tau$, which plays a critical role in mining hard negatives \cite{yu2023self}. As shown in Fig.~\ref{fig7}(a), when $\tau=0.2$, LSIDN performs the best, whereas smaller or larger values will result in poor performance; (\rmnum 2) the contrastive loss weight $\lambda$, which controls the strength of contrastive regularization, too large $\lambda$ will conflict with the main task and too small $\lambda$ will lead to ineffective regularization. The results shown in Fig.~\ref{fig7}(b) support the view of point, where the optimal value of $\lambda$ is 0.1; (\rmnum 3) the session division threshold $\omega$, which controls the division criteria of the user behavior sequence, small $\omega$ leads to over-segmentation while large $\omega$ leads to under-segmentation, both of which make segmented sessions sub-optimal for representing session-level interests. As shown in Fig.~\ref{fig7}(c), 360 minutes is the best for $\omega$, while smaller or larger values both impair performance. Note that we keep all other hyperparameters optimal when investigating each hyperparameter.

 \section{Conclusion}

In this paper, we propose a long- and short-term interest denoising network, which employs different encoders and tailored denoising strategies to extract different time-scale user interests, respectively, thus achieve more comprehensive and robust user modeling. Specifically, we employ a session-level interest extraction and evolution strategy to avoid introducing inter-session behavioral noise into long-term interest modeling; we also resort to contrastive learning equipped with a novel augmentation method tailored for short sequences to alleviate the impact of unintentional behavioral noise on short-term interest modeling. Extensive results show that LSIDN consistently outperforms state-of-the-art models and achieves significant robustness. 
\section*{Acknowledgement}
This work was supported by the National Natural Science Foundation of China under Grant No. 62072450 and Meituan.

\end{document}